# NaSn$_2$As$_2$: An Exfoliatable Layered van der Waals Zintl Phase

*Maxx Q. Arguilla,[‡,1] Jyoti Katoch,[‡,2] Kevin Krymowski,[3] Nicholas D. Cultrara,[1] Jinsong Xu,[2] Xiaoxiang Xi,[4] Amanda Hanks,[3] Shishi Jiang,[1] Richard D. Ross,[1] Roland J. Koch,[5] Søren Ulstrup,[5] Aaron Bostwick,[5] Chris Jozwiak,[5] Dave McComb,[3] Eli Rotenberg,[5] Jie Shan,[4] Wolfgang Windl,[3] Roland K. Kawakami[2] and Joshua E. Goldberger\*,[1]*

[1]Department of Chemistry and Biochemistry, The Ohio State University, Columbus, Ohio 43210-1340, United States

[2]Department of Physics, The Ohio State University, Columbus, Ohio 43210-1340, United States

[3]Department of Materials Science and Engineering, The Ohio State University, Columbus, Ohio 43210-1340, United States

[4]Department of Physics, The Pennsylvania State University, University Park, Pennsylvania 16802-6300, United States

[5]Advanced Light Source, E. O. Lawrence Berkeley National Laboratory, Berkeley, California 94720, United States

**Abstract:** The discovery of new families of exfoliatable 2D crystals that have diverse sets of electronic, optical, and spin-orbit coupling properties, enables the realization of unique physical phenomena in these few-atom thick building blocks and in proximity to other materials. Herein, using $NaSn_2As_2$ as a model system, we demonstrate that layered Zintl phases having the stoichiometry $ATt_2Pn_2$ (A = Group 1 or 2 element, Tt = Group 14 tetrel element and Pn = Group 15 pnictogen element) and feature networks separated by van der Waals gaps can be readily exfoliated with both mechanical and liquid-phase methods. We identified the symmetries of the Raman active modes of the bulk crystals *via* polarized Raman spectroscopy. The bulk and mechanically exfoliated $NaSn_2As_2$ samples are resistant towards oxidation, with only the top surface oxidizing in ambient conditions over a couple of days, while the liquid-exfoliated samples oxidize much more quickly in ambient conditions. Employing angle-resolved photoemission spectroscopy (ARPES), density functional theory (DFT), and transport on bulk and exfoliated samples, we show that $NaSn_2As_2$ is a highly conducting 2D semimetal, with resistivities on the order of $10^{-6}$ Ω m. Due to peculiarities in the band structure, the dominating p-type carriers at low temperature are nearly compensated by the opening of n-type conduction channels as temperature increases. This work further expands the family of exfoliatable 2D materials to layered van der Waals Zintl phases, opening up opportunities in electronics and spintronics.

Over the past few years there have been numerous exciting physical phenomena that have been both discovered and theoretically predicted to occur in single and few-layer thick van der Waals (vdW) materials.[1-7] The initial discoveries of Dirac physics in graphene,[2,3,8] and thickness-dependent tunability of band gaps in transition metal dichalcogenides (TMDs)[6,9-12] have led to an explosion of exotic phenomena in single-layer and few-layer materials such as the valley Hall effect,[13-15] magnetic proximity effects,[16-19] and topological insulator phases.[1,20,21] The existence of such a plethora of new phenomena in ultrathin layers stems from their sensitivity to doping,[22-27] local chemical environments,[28,29] electrical gating,[6] as well as the elimination of inversion symmetry to access spin-based phenomena.[13,30,31] Consequently, there is considerable interest in these materials for applications ranging from opto/spin/electronics[13,17,32-38] to transparent conductors[24,39,40] to sensing[41-44] to catalysis.[45-47]

To date most of these reports have focused on graphenes,[2,16,17,21-24,41] group 14 graphene and graphane analogues,[48-51] transition metal MXenes,[52-54] and TMDs.[6,9,13,14,34,35,42,43,55-58] A much less explored class of potential vdW materials are the layered Zintl phases.[59-65] Layered Zintl phases are typically comprised of layers of Group I or Group II cations that are ionically bonded to layers of $p$-block elements that form a covalently bonded extended lattice. One potential advantage of building two dimensional (2D) materials from main group elements compared to transition metals, is the higher conductivities and mobilities that can be accessed as a result of the conduction and valence bands being comprised of *p*-orbitals, which have much broader band dispersion than *d*-orbitals.  Second, many of these materials are shown to be metals,[66] semiconductors,[67] and even superconductors (*i.e.* BaGaSn).[63] There have also been numerous predictions of exotic physics, such as topological nodal-line characters in $ATt_2$ phases (A = Ca, Sr and Ba, Tt = Si, Ge and Sn)[68] and 3D-Dirac states in $SrSn_2As_2$.[69] Also, the large spin-orbit coupling inherent to the heavy atoms in the framework of these materials can give rise to various spintronic phenomena, especially when combined with perpendicular electric fields that can be maximized upon exfoliation.[70] Furthermore, they have been investigated as hydrogen storage materials such as in the layered polyanionic hydrides (*i.e.* $SrAl_2 \rightarrow SrAl_2H_2$).[71,72] Finally, extensive work by our lab has shown that layered Zintl phases can be readily transformed into hydrogen- and organic-functionalized graphane materials.[48-51,73]  Being able to create exfoliatable Zintl phase derivatives will enable future studies on the layer dependence of these electronic, topological, and spintronic properties and may lead to exciting phenomena.

There are numerous examples of layered Zintl phases that have weak bonding between the layers, which can enable their exfoliation.[59,60] These Zintl phases have the general formula of $ATt_2Pn_2$ (A = Group 1 or 2, Tt = Group 14 tetrel element and Pn = Group 15 pnictogen element), of which there have been reported at least 7 different phases and alloys.[59,60] These materials consist of the Group I/II element being sandwiched between two layers of honeycomb networks of puckered alternating BN-like TtPn layers. Due to the greater electronegativity, the pnictogens are tilted to be in close proximity to the Group I/II element, and the Tt atom is oriented towards the Tt atoms in an adjacent layer (**Figure 1a**). From electron-counting arguments, there is weak-to-no covalent bonding between Tt atoms in neighboring TtPnAPnTt layers. In $NaSn_2As_2$, each As atom, which is in close proximity to the Na atom, has 3 covalent bonds and formally 8 electrons, while each Sn atom has 3 covalent bonds and thus 7.5 electrons. In $SrSn_2As_2$, each Sn atom would formally have 8 electrons. Therefore, the weak bonding between the Sn atoms in these phases could enable these materials to be readily exfoliated. To date, there have been little to no experimental studies about the vibrational and electronic behavior of these materials, their air-sensitivity, and no studies on their exfoliation.

Herein, we show that layered vdW Zintl phases can be readily exfoliated using both mechanical and liquid-phase processes, using $NaSn_2As_2$ as a model system. We also rigorously characterize the vibrational and electronic properties and air stability of $NaSn_2As_2$ crystals. We evaluate the bulk Raman spectrum and identify the peak symmetries of the modes *via* polarized Raman spectroscopy. We show that $NaSn_2As_2$ is resistant towards oxidation, with surface oxidization occurring in ambient conditions over a couple of days. Finally, using a combination of angle-resolved photoemission spectroscopy (ARPES), density functional theory (DFT), and bulk and layer-dependent transport, we show that this material is an excellent 2D semimetal, with resistivities on the order of $10^{-6}$ Ω m, both in bulk and when exfoliated to few layers.

**Results and Discussion**

NaSn$_2$As$_2$ crystallizes in a rhombohedral R-3m unit cell characterized by three centrosymmetric anionic bilayers of Sn and As held by six-coordinated Na cations (**Figure 1a**). These bilayers (SnAsNaAsSn) are bound by vdW forces with an approximate 3.3 Å gap between the Sn atoms of adjacent bilayers (**Figure 1a**). This vdW gap that exists between the bilayers allows for the possible exfoliation of NaSn$_2$As$_2$ down to SnAsNaAsSn bilayers, which from the crystal structure corresponds to ~9.2 Å per bilayer. Bulk crystals of NaSn$_2$As$_2$ were grown *via* a tube melt synthesis of elemental precursors of Na, Sn and As with a 1:2:2 stoichiometric loading. The resulting mm- to cm-sized crystals exhibit a metallic luster and a c-axis faceting (**Figure 1a**, bottom right). X-ray diffraction (XRD) measurements on NaSn$_2$As$_2$ powders followed by Rietveld refinement (**Figure S1**) reveal that the as-grown crystals are phase pure with an a-lattice parameter of 3.99998(10) Å and a c-lattice parameter of 27.5619(13) Å, consistent with previous reports (**Figure 1b**).[59] We also confirm the 1:1 stoichiometry between the Sn and As atoms *via* X-Ray Fluorescence (XRF) measurements (**Figure S2**).

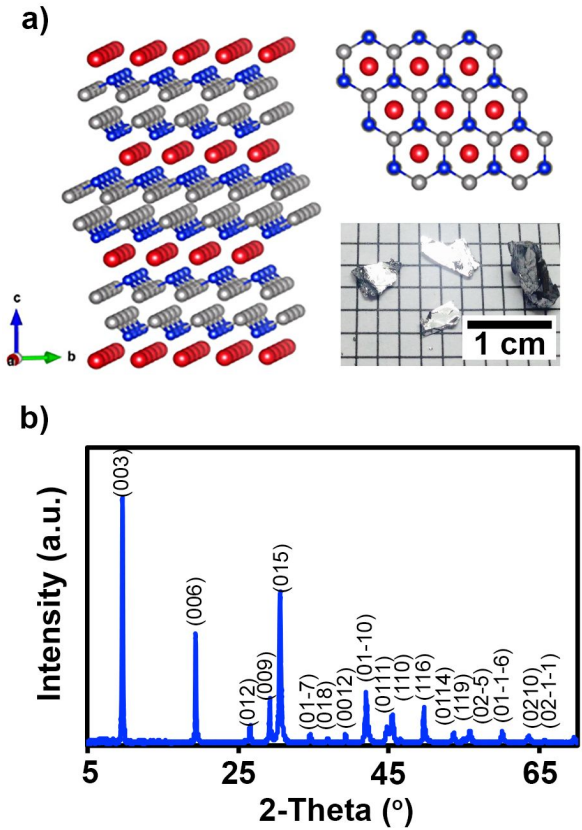

**Figure 1.** (a) Crystal Structure of $NaSn_2As_2$ (Na, red; Sn, gray; As, blue). Top right: $NaSn_2As_2$ crystal structure down the *c*-axis. Bottom right: mm- to cm-size $NaSn_2As_2$ crystals. (b) Powder-XRD of $NaSn_2As_2$. Enclosed in parentheses are the rhombohedral miller indices.

The Raman spectrum of bulk $NaSn_2As_2$ shows three peaks having narrow full-width-at-half-maximum (FWHM). These Raman stretches occur at ~182 cm$^{-1}$, ~206 cm$^{-1}$ and ~228 cm$^{-1}$ with FWHM of 5 cm$^{-1}$, 3 cm$^{-1}$, and 6 cm$^{-1}$, respectively (**Figure 2a**). In the R-3m space group ($D_{3d}$ point group), these Raman modes can correspond to either $A_{1g}$ or $E_g$ symmetry. According to group theory, there should be 15 total vibrational branches at the Γ point in the Brillouin zone, of which only 6 are Raman active (2 $E_g$ + 2 $A_{1g}$). To determine the symmetry of each Raman mode, we performed polarized (0° to 360°) backscattered Raman measurements along the crystalline *c*-axis (**Figure 2b,c**). According to the selection rules for this space group, both the $A_{1g}$ and $E_g$ modes should appear in the co-polarized–Z(XX)Z geometry, while only the $E_g$ mode is allowed in cross-polarized –Z(XY)Z geometry. There is very little difference in the relative intensity of these Raman modes in the co-polarized vs. non-polarized configurations. However, in the cross-polarized geometry, the peak centered at ~228 $cm^{-1}$ disappears, and the peak centered at ~206 $cm^{-1}$ significantly decreases in intensity relative to the ~182 $cm^{-1}$ peak. The complete disappearance of the peak centered at 228 $cm^{-1}$ indicates that it is an $A_{1g}$ mode. The decrease in the intensity of the Raman mode at 206 $cm^{-1}$ suggests that it is comprised of at least an $A_{1g}$ and an $E_g$ Raman modes that are close in energy. Finally, the lack of polarization dependence in the 182 $cm^{-1}$ mode suggests that it is of $E_g$ origin.

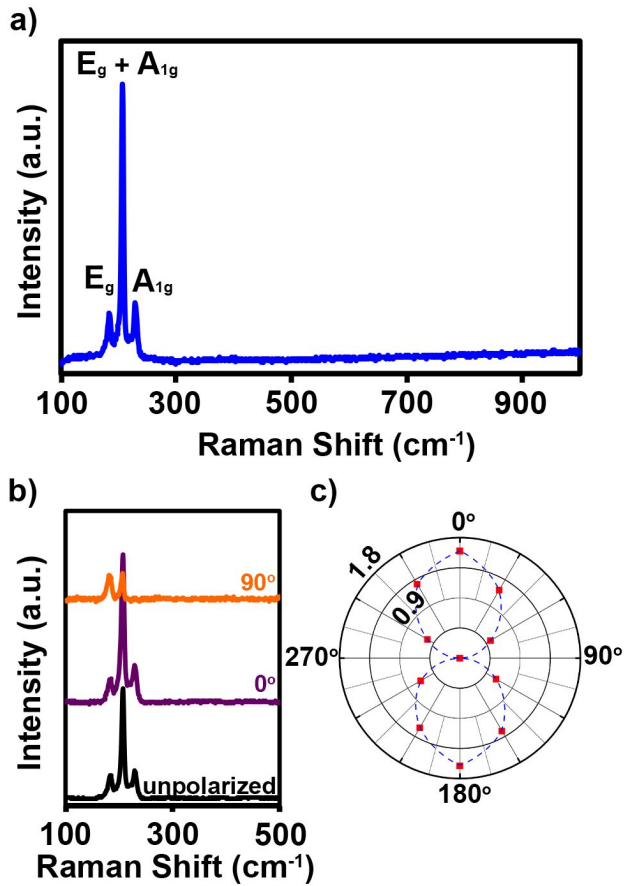

**Figure 2.** (a) Raman spectrum of NaSn$_2$As$_2$. (b) Polarized (at 0° and 90°) and unpolarized Raman spectra of NaSn$_2$As$_2$. (c) Radial plot (0° to 360°) of the A$_{1g}$/E$_{2g}$ intensity ratios at 228 cm$^{-1}$ and 182 cm$^{-1}$, respectively, from polarized Raman experiments. Blue dashed trace represents the fitted curve.

X-ray photoelectron spectroscopy (XPS) measurements on bulk flakes are also indicative of anionic [SnAs]$^-$ layers (**Figure 3**). From the binding energy of the Na *1s* peak (1071 eV), it can be deduced that the Na is cationic in nature with its energy close to a Na$^+$ binding energy.[74] Additionally, Sn is anionic based on the binding energy of the Sn $3d_{5/2}$ peak (484 eV) which is significantly lower than the $Sn^0$ $3d_{5/2}$ binding energy (485.2 eV).[75] Comparing the As $3d_{5/2}$ peak binding energy (39.8 eV) of $NaSn_2As_2$ to As $3d_{5/2}$ peaks of GaAs (41.1 eV) and As (42.0 eV) reference standards indicated a more anionic As species than in GaAs (**Figure 3d**). All together, these oxidation states indicate an anionic 2D [SnAs]⁻ lattice which is charge-balanced by $Na^+$.

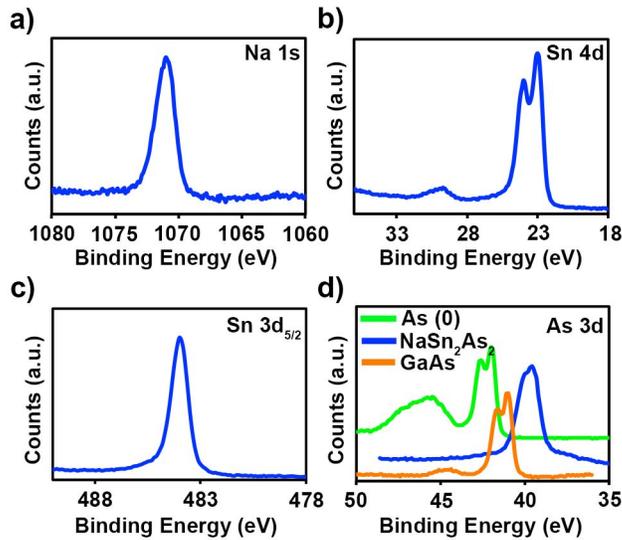

**Figure 3.** XPS of $NaSn_2As_2$ showing the Na 1s, Sn 4d Sn $3d_{5/2}$ and As 3d peaks.

The air stability of these crystals was first studied by exposing ground powders and mechanically exfoliated flakes (~50-100 nm) in ambient air for 7 days. From atomic force microscopy (AFM) measurements, there is minimal change in surface roughness of the flakes following air exposure from 0.39 nm to 0.46 nm (**Figure 4a**). XRD measurements on the air-exposed samples show no new oxidized phases nor significant broadening of the diffraction peaks (**Figure 4b**). From the Raman analyses of ~100 nm flakes, there were no new vibrational modes or features (which might indicate oxidation) and no significant increase in the FWHM of the peaks corresponding to the lattice vibrational modes before and after 1 week of air exposure (**Figure 4c**). No asymmetric vibrational modes corresponding to either Sn-O (500-700 cm$^{-1}$)[76,77] or As-O (700-900 cm$^{-1}$)[78] bond formation were observed from Fourier-transform infrared (FTIR) measurements of pristine and air-exposed powders (**Figure 4d**). XPS measurements show the oxidation of the Sn and As atoms on surface of NaSn$_2$As$_2$ after one week of exposure through the appearance of a shoulder centered around 485.7 eV (accounting for 26.5% and indicative of the presence of a Sn$^{4+}$ *3d$_{5/2}$* peak) and around 42.8 eV (accounting for 11.5% and indicative of the presence of a As$^{3+}$ *3d$_{5/2}$* peak), respectively (**Figure S3**). These peaks disappear after Ar ion etching the top 1 nm (~1 bilayer), calibrated for SiO$_2$. These measurements suggest that the bulk of NaSn$_2$As$_2$ is air stable with only the top surface bilayer oxidizing over time.

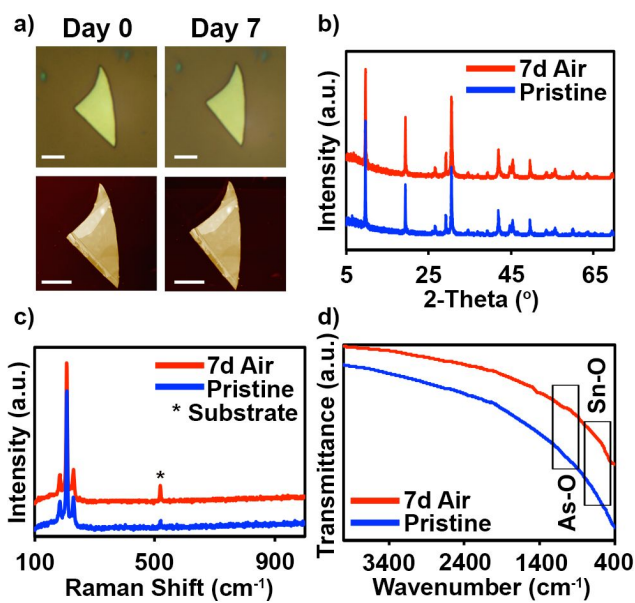

**Figure 4.** (a) Optical and AFM micrographs (Thickness: 65 nm; Scale bars correspond to 4 μm), (b) Powder XRD, (c) Raman spectra and (d) FTIR spectra of NaSn$_2$As$_2$ before and after exposure to air for 7 days.

To demonstrate the possibility of exfoliating NaSn$_2$As$_2$ into few- to single-bilayers, bulk single crystals were exfoliated onto 285 nm SiO$_2$/Si substrates *via* mechanical exfoliation using Scotch™ tape. From this technique, thin layers with thicknesses down to 3.6 nm, which corresponds to approximately four SnAsNaAsSn bilayers, or 4/3 of a NaSn$_2$As$_2$ unit cell (*c* lattice parameter = 27.56 Å), were observed (**Figure 5a, S4**). Mechanical exfoliation produced numerous few-layer flakes with multiple steps (**Figure S5**), with step thicknesses that were multiples of ~0.9 ± 0.2 nm (**Figure S6**). The fact that these step heights only varied by this multiple, show that the cleavage occurs either between the Sn•••Sn planes or the Na•••As planes, but not both. To determine which plane will more likely cleave upon exfoliation, we calculated the adhesion energies by determining the difference in overall energy between the bulk structure and both NaAsSnSnAs layers and SnAsNaAsSn separated by 10 Å vacuum, using the DFT-D2 method by Grimme[79] which includes both chemical interactions from a generalized gradient approximation (GGA) and a dispersive term which describes vdW interactions. The adhesion energy between neighboring Sn-Sn layers was calculated to be 1.45 J/m$^2$, and 1.75 J/m$^2$ between neighboring Na•••As layers. This validates the assignment of NaSn$_2$As$_2$ exfoliating into SnAsNaAsSn layers. Future scanning tunneling microscopy measurements would provide further verification of the surface termination of these exfoliated flakes.

Aside from mechanical exfoliation, layered 2D materials can also be exfoliated through liquid-phase exfoliation when sonicated using a suitable solvent. Here, NaSn$_2$As$_2$ powders were dispersed in various solvents from isopropyl alcohol (low surface tension) to dimethyl sulfoxide (high surface tension) using sonication (**Figure 5b**). From these solvents, absorbance measurements revealed that the dispersion in N-cyclohexyl-2-pyrrolidone (CHP) yielded the highest resulting concentration of the supernatant after centrifugation (**Figure S7**). The Raman spectra show that the exfoliated flakes still maintain the lattice vibrational modes, which indicates that the structure is still intact after exfoliation (**Figure 5c**). The thicknesses of the dispersed NaSn$_2$As$_2$ layers were analyzed by taking AFM micrographs of the drop-casted suspension onto 285 nm

SiO$_2$/Si substrates (**Figure 5d, S8**). The thicknesses of these dropcast flakes ranged from 1 to 9 nm, corresponding to 1 to 9 SnAsNaAsSn bilayers. Additionally, the lengths and widths of these dropcast flakes ranged from 200-500 nm. The average thickness from a histogram of 36 different flakes was 4.6 nm (**Figure S9**). To further probe the structure and morphology of the liquid-exfoliated NaSn$_2$As$_2$ flakes, we performed transmission electron microscopy (TEM) on filtered NaSn$_2$As$_2$ dispersions on lacy carbon grids. In these measurements, we were able to observe bright-field TEM micrographs with thin sheets of NaSn$_2$As$_2$ that have less contrast than the 10 nm lacey carbon grid, and with lengths and widths on the order of ~100 nm (**Figure 5e**). A typical electron diffraction pattern from one of these NaSn$_2$As$_2$ flakes can be indexed to a simple hexagonal unit cell with $a = b \approx 4.02$ Å, assuming a [001] zone axis. This is in close agreement with the X-ray crystal structure. Further confirming the NaSn$_2$As$_2$ composition, the energy-dispersive X-ray spectrum (EDX) has strong Na, Sn, and As signal (**Figure S10**). Additionally, crystalline domain sizes that are greater than 50 nm have been observed in high-resolution transmission electron microscopy (HRTEM) (**Figure S11**). XPS measurements show that oxidation of these very thin, solution-exfoliated samples, occurs much more readily, even after a day of air exposure (**Figure S12**). Whether oxidation occurs due to the reduced layer thickness, lateral size, or *via* sonication induced defects is subject to future studies. Still, similar to almost all other 2D vdW materials, encapsulation methods and surface functionalization will be required in order to achieve long-term air stability from these solution-exfoliated few-layer samples.[80-83]

As discussed previously, formal electron counting would suggest that $NaSn_2As_2$ would have 0.5 bonds between neighboring Sn-Sn atoms. In order to determine the exact nature of this Sn•••Sn interaction, we calculated the energy of the $NaSn_2As_2$ bulk hexagonal cell, which contains three SnAsNaAsSn layers, as a function of interlayer spacing up to 10 Å separation. Two calculations were performed: One with the DFT-D2 method by Grimme[79] that combines GGA with vdW interactions, and the other with straight GGA, which has been shown for the case of graphite to include no vdW interactions.[84] The results are shown in **Figure S13.** While the binding energy between the layers without vdW forces is 0.88 $J/m^2$, inclusion of vdW forces adds 0.57 $J/m^2$ for a total of 1.45 $J/m^2$. These values are larger than the adhesion energy of other 2D vdW, which typically range from 0.2-0.6 $J/m^2$.[83,85,86] Still, the vdW interaction makes up 40% of the interaction energy and thus has nearly equal strength to the orbital interaction between the Sn layers. This significant vdW component further justifies the classification of this material as a vdW phase.

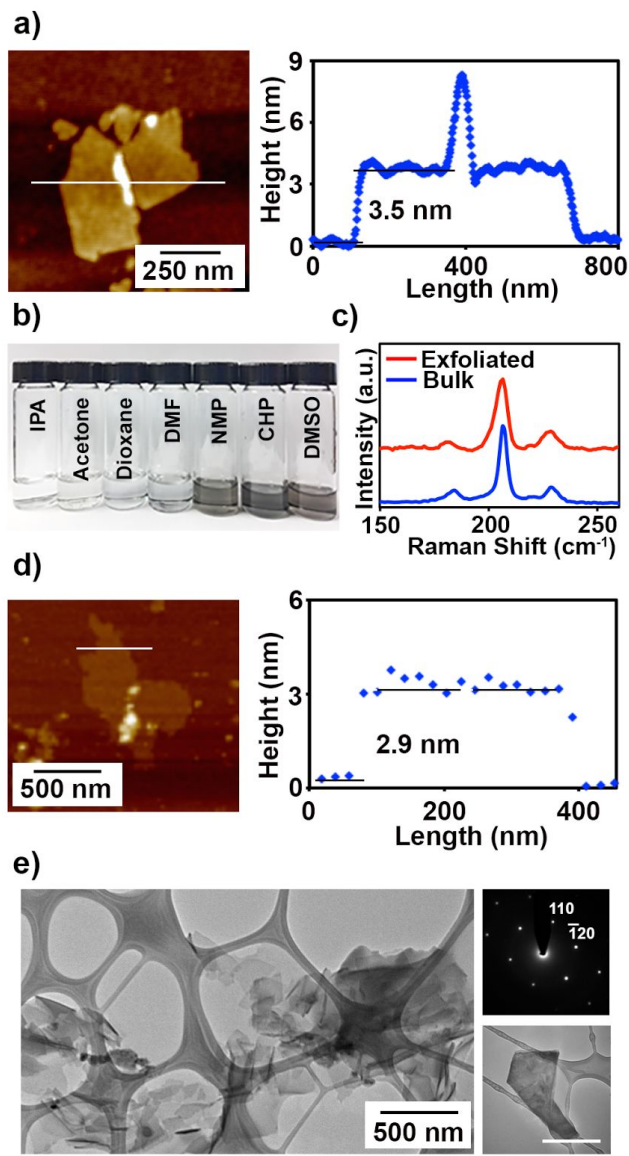

**Figure 5.** (a) AFM image of few-layers of NaSn$_2$As$_2$ from mechanical exfoliation with the corresponding thickness profile. (b) NaSn$_2$As$_2$ dispersions in solvents with increasing surface tension from left to right (IPA = isopropyl alcohol, DMF = dimethylformamide, NMP = N-methyl-2-pyrrolidone, CHP = N-cyclohexyl-2-pyrrolidone and DMSO = dimethyl sulfoxide). (c) Raman of drop-casted few-layer (average thickness = 4.6 nm)

$NaSn_2As_2$ in 285 nm $SiO_2$/Si substrate. (d) AFM image of liquid-exfoliated $NaSn_2As_2$ with the corresponding thickness profile. (e) TEM of liquid-exfoliated few layers of $NaSn_2As_2$ with a representative selected area electron diffraction (SAED) pattern, scale bar is 500 nm.

ARPES measurements were collected to elucidate the electronic structure of $NaSn_2As_2$ (**Figure 6a, S14**), and compared to high-level DFT calculations with SOC (**Figure 6b**). We show data at photon energy 127 eV, but variation of the photon energy showed little or no change in the observed bands near $E_F$ (apart from intensity due to optical matrix element variation), indicating a quasi-2D character. There is excellent agreement between experiment and theory where the only discrepancy is a shallow electron pocket observed around Γ observed by ARPES, which was predicted to be unoccupied in the DFT calculation. The atomic and orbital contributions of each band were calculated and are shown in Figure 6c. The bands localized at -1 eV close to Γ point correspond to filled As *p*-bands. The partially unfilled band that the Fermi level cuts through primarily consists of Sn-*s* and Sn-*p* character, with some As-*p* contributions. The greater filling of the As bands compared to the Sn bands is consistent with the greater electronegativity of As.

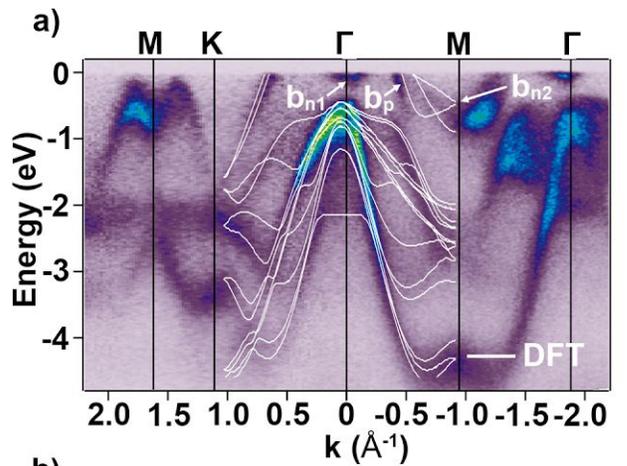
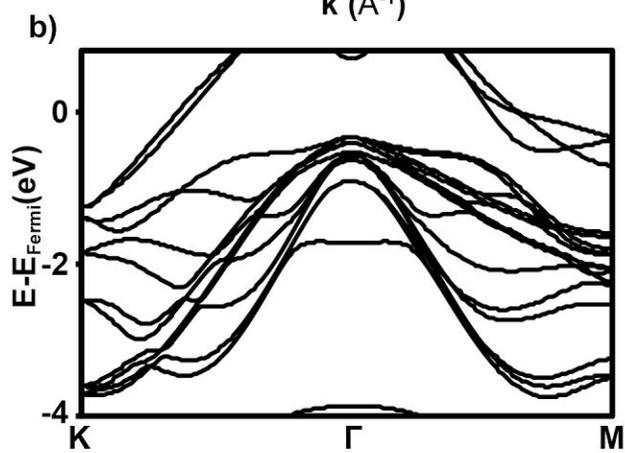
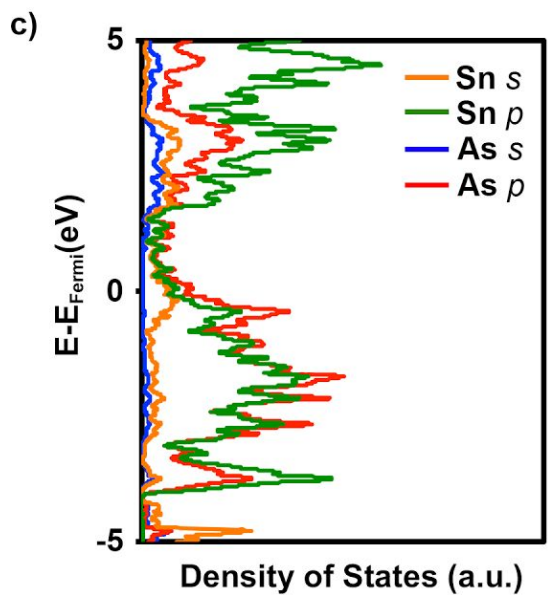

**Figure 6**, Electronic Band Structure of $NaSn_2As_2$. (a) Overlay of the theoretical band structure with the ARPES spectrum of a cleaved $NaSn_2As_2$ single crystal, acquired at photon energy 127 eV and T = 41 K. The bands that correspond to the electron pockets are labeled as $b_{n1}$ and $b_{n2}$, while the bands that comprise the hole pockets are labeled as $b_p$. (b) Heyd-Scuseria-Ernzerhof (HSE) band structure of $NaSn_2As_2$ with spin-orbit coupling (SOC). (c) Density of states of the contributing orbitals in $NaSn_2As_2$ near the Fermi level.

Two-probe temperature-dependent resistivity measurements were performed on $NaSn_2As_2$ single crystals to further elucidate the electronic structure. At room temperature the in-plane resistivity of a representative $NaSn_2As_2$ flake was determined to be 2.9 $\mu\Omega \cdot m$, which is highly conducting. The resistivity decreases with decreasing temperatures, which is consistent with metallic behavior as a consequence of phonon-mediated carrier scattering at higher temperatures (**Figure 7a**). Longitudinal and transverse resistivity measurements have also been performed on an exfoliated flake (thickness t = 350 nm) patterned into a Hall bar device. The room temperature in-plane resistivity is 1.7 $\mu\Omega \cdot m$, which is comparable with that in the bulk. The temperature dependent longitudinal resistivity again reveals the metallic nature of $NaSn_2As_2$, with a similar residual resistivity ratio as observed in the bulk. The transverse resistivity was measured under magnetic fields applied perpendicularly to the basal plane at 2.1, 100, 200, and 300 K. The Hall coefficient is positive over this temperature range, suggesting p-type conduction, although it approaches zero at room temperature.

The low-temperature p-type character, and the observed compensation at room temperature is consistent with the band structure measured by ARPES and calculated by DFT. To understand the nature of the charge carriers, we refer to the ARPES measurement (**Figure 6a**). Here, a p-type band, $b_p$, crosses the $E_f$. Also, a small central electron pocket is observed at $\Gamma$, labeled as $b_{n1}$ and, just below $E_f$, another band ($b_{n2}$) forms a closed electron pocket ~0.75 eV deep centered on the M point. At low temperatures, these bands do not contribute carriers due to the low intersection at $E_f$, predicting a predominantly hole (p-type) carrier density. Thermal smearing of the Fermi level at RT excites carriers into the electron-like pocket around the M-point, dramatically increasing the ratio of n-type to p-type carriers. A detailed discussion of this phenomenon (temperature-dependent carrier type) is beyond the scope of this manuscript and will be elaborated upon by temperature-dependent ARPES and thermopower measurements in a subsequent study. Explorations of the layer dependence of this temperature-dependent multiband transport can potentially lead to interesting phenomena and DFT simulations of the electronic structure of an SnAsNaAsSn layer isolated by 10 Å vacuum predicts that the metallic behavior is retained in a single layer (**Figure S15**).

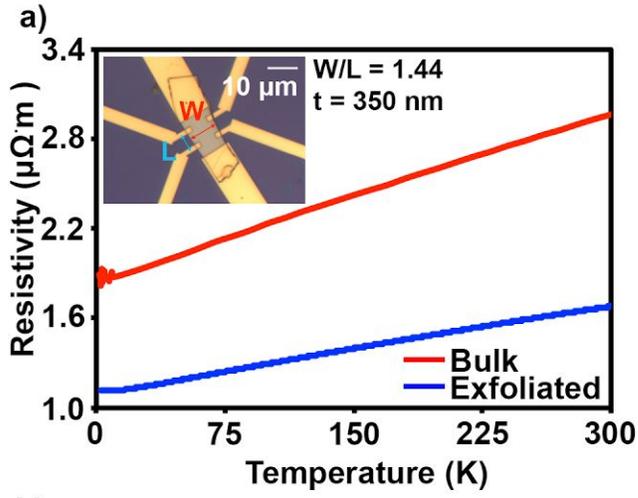

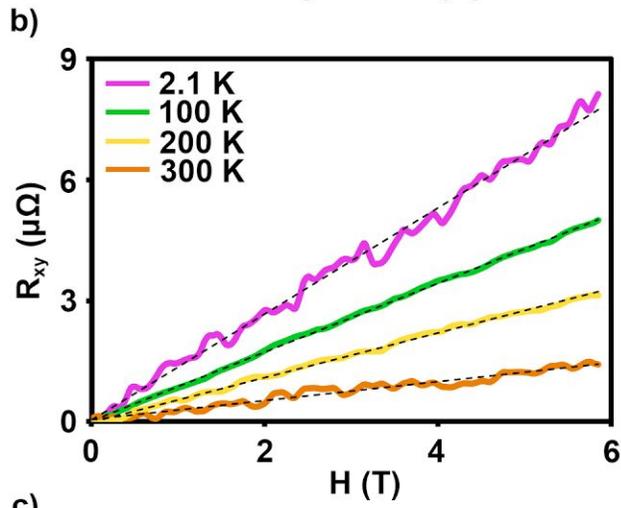

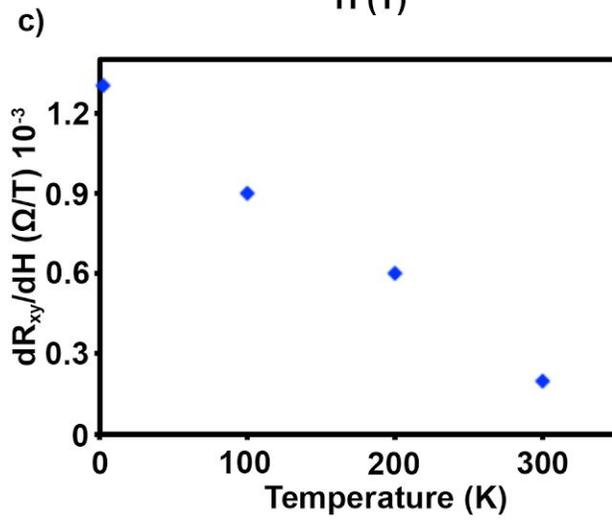

**Figure 7.** (a) Temperature-dependent in-plane resistivity of bulk and exfoliated $NaSn_2As_2$ down to 2.1 K. Inset is an optical microscope image of the Hall bar device. The thickness of the flake is 350 nm and the width to length ratio of the channel is 1.44. (b) The magnetic field dependence of the transverse resistance of the Hall bar device at 4 different temperatures. (c) The Hall coefficient extracted from (b).

**Conclusion**

In summary, we have demonstrated that layered vdW Zintl phases such as $NaSn_2As_2$ can be exfoliated into single- and few-layer (SnAsNaAsSn) sheets. From theory, ARPES and electronic transport, these bulk crystals are metallic in nature, but the unique band features allow for varying degrees of two-carrier conduction at various temperatures. At low temperatures p-type carriers dominate, but at higher temperatures n-type conduction channels become accessible. This work opens up the future exploration of layer-dependent phenomena of $NaSn_2As_2$ and other numerous materials in this family of high spin-orbit coupling vdW Zintl phases for applications ranging from transparent conductors to spintronics.

**Experimental**

**Growth.** Bulk crystals of $NaSn_2As_2$ were grown *via* a quartz tube melt synthesis under a ~50 mTorr pressure with 1:2:2 Na:Sn:As loading. The tubes were heated to 750ºC for 3 to 5 hours, cooled to room temperature for 40 to 60 hours followed by a 50-hour annealing at 400ºC. The preparation and handling of all the reagents were done in an Ar-filled glovebox. Sodium metal cubes (Na, Sigma Aldrich, 99.9%) were used after the oxide layer was stripped off *via* mechanical cleavage. Tin (Sn, STREM Chemicals, 99.8%) and arsenic (As, STREM Chemicals, 99%) were purchased and used without further purification. Note that since As is toxic, extreme caution should be observed when handling this compound.

**Characterization.** The crystal structure and phase purity of the as-grown crystals were identified *via* flat plane X-Ray diffraction measurements done on a Bruker D8 powder X-Ray diffractometer (sealed Cu X-Ray tube: 40 kV and 50 mA). The Raman scattering spectra were collected using a Renishaw InVia Raman equipped with a Pike Technologies KRS-5 polarizer and a charge-coupled device detector. The Raman spectra were collected using a 633 nm laser as an illumination source. Fourier-Transform Infrared measurements were performed using a Perkin-Elmer Frontier Dual-Range Far-IR/Mid-IR spectrometer. The stoichiometric elemental ratio was confirmed *via* X-Ray fluorescence analysis using an Olympus DELTA handheld X-Ray fluorimeter calibrated with Sn and As standards of varying ratios. The X-ray photoelectron spectra of the samples were taken using a Kratos Axis Ultra X-ray photoelectron spectrometer with a monochromated Al X-ray source. Energy calibration was performed using the C *1s* peak.

**Air Stability.** $NaSn_2As_2$ powders and flakes exfoliated onto 285 nm $SiO_2$/Si substrates were exposed to ambient air for different durations. The resulting powders/flakes were analyzed for signs of oxidation *via* atomic force microscopy using a Bruker AXS Dimension Icon Atomic Force Microscope, FTIR and Raman spectroscopy, XRD and XPS.

**Mechanical Exfoliation.**

The $NaSn_2As_2$ crystal was exfoliated onto 300 nm or 285 nm $SiO_2$/Si substrates. The substrates were cleaned by sonication for 5 minutes each in acetone and isopropyl alcohol. The $NaSn_2As_2$ crystals were micro-mechanically cleaved using a silicone-free tape until the tape had uniform coverage of thin $NaSn_2As_2$ flakes. Then this tape is pressed facing down on the cleaned wafers to obtain thin exfoliated flakes on the substrate.

**Liquid-Phase Exfoliation.** Ground samples of $NaSn_2As_2$ (5 mg/mL) were sonicated for 4 hours using various solvents: IPA, acetone, 1,4-dioxane, dimethylformamide, N-methyl-2-pyrrolidone, N-cyclohexyl-2-pyrrolidone and dimethyl sulfoxide. To separate the non-dispersed flakes, the dispersions were centrifuged at 1000 rpm for 90 minutes. The supernatant was collected and a part of it was further centrifuged at 3000 rpm for another 90 minutes. To determine the best solvent for exfoliation, absorbance measurements of the dispersion were taken on a Perkin-Elmer Lambda 950 UV/Vis/NIR spectrophotometer. For AFM, Transmission Electron Microscopy measurements and high-resolution transmission electron microscopy (HRTEM), dispersions in CHP were drop-casted onto 285 nm $SiO_2$/Si substrates and 200 mesh lacy carbon Cu grids. The substrates and grids were dried for 4 to 12 h under vacuum at ambient temperature. To remove any residual CHP, the substrates for AFM were further rinsed (3x) with IPA. The Selected Area Electron Diffraction and EDX spectroscopy were performed in an FEI/Philips CM-200T and CM-12T TEM for phase identification. HRTEM was performed using an FEI Image-Corrected Titan3 G2 60-300 S/TEM at 300 keV. All solvent and sample transfers, decantation and sample preparation were done in an Ar-filled glovebox. The samples, on the other hand, were kept in Ar-filled sealed vials during the sonication and centrifugation steps.

**ARPES and DFT Calculations.**

The ARPES measurements were performed in the micro-ARPES chamber of the MAESTRO beamline at the Advanced Light Source. The chamber is equipped with a Scienta R4000 hemispherical electron energy analyzer. We successfully obtained ARPES spectra on high quality $NaSn_2As_2$ single crystals, which were cleaved *in-situ*. The measurements were done with photon energy $h\nu$ = 127 eV, with an overall energy resolution of about 30 meV. The sample was cooled down to 41 K with liquid helium, and the base pressure in the chamber during the measurements was better than $2 \times 10^{-11}$

Torr. The electronic band structure was calculated with density functional theory based on a Heyd-Scuseria-Ernzerhof hybrid functional method using the Vienna Ab-initio Simulation Package (VASP).[33-38] Due to the rhombohedral nature of the $NaSn_2As_2$ unit cell, we expanded the unit cell to a hexagonal cell, which contains three times the number of atoms as the rhombohedral unit cell. The mixing parameter, $\alpha$, was set to 0.25 in these calculations. To further accommodate the exchange effects, we included spin orbit coupling (SOC) in our calculations.

**Electronic Transport.** Two-probe temperature-dependent resistivity measurements were performed on a $NaSn_2As_2$ single crystal with pressed indium contacts using a Quantum Design 14 T Physical Properties Measurement System (PPMS) from 2 K to 300 K. For few-layer devices, $NaSn_2As_2$ flakes were mechanically exfoliated from bulk crystals onto $SiO_2$/Si substrates using Scotch tape. Their thickness was determined by AFM measurements. Flat flakes with good geometry were chosen to make Hall bar devices by depositing Ti/Au electrodes using standard electron beam lithography and evaporation techniques. The longitudinal and transverse resistance values were also measured in a PPMS down to 2 K, with the excitation current limited to 10 μA. To account for magnetoresistance and longitudinal-transverse coupling, we anti-symmetrized the transverse resistance $R_t$ under opposite magnetic field directions to obtain the Hall resistance $R_{xy}$ (H).

ASSOCIATED CONTENT

**Supporting Information**. Powder XRD Rietveld refinement results, XRF calibration curve and Sn:As ratio determination, Air exposure-dependent XPS spectra of Na 1s, Sn 4d, Sn $3d_{5/2}$ and As 3d peaks of $NaSn_2As_2$, AFM images of mechanically-exfoliated $NaSn_2As_2$, AFM step-height histogram of mechanically-exfoliated $NaSn_2As_2$, UV-Vis absorption spectrum of $NaSn_2As_2$ dispersed in various solvents, CHP-exfoliated $NaSn_2As_2$ AFM images and histogram, EDX spectrum, HRTEM image of CHP-exfoliated $NaSn_2As_2$, Air exposure-dependent XPS spectra of Na 1s, Sn $3d_{5/2}$ and As 3d peaks of CHP-exfoliated $NaSn_2As_2$, adhesion energy of $NaSn_2As_2$ as a function of layer separation, ARPES spectrum of $NaSn_2As_2$ and HSE single-layer electronic band structure of $NaSn_2As_2$. This material is available free of charge *via* the Internet at http://pubs.acs.org.

AUTHOR INFORMATION

**Corresponding Author**

*Goldberger@chemistry.ohio-state.edu

**Author Contributions**

[‡]These authors contributed equally.

**Funding Sources**

Funding for this research was provided by the Center for Emergent Materials: an NSF MRSEC under award number DMR-1420451. Partial funding for this research was provided by NSF EFRI-1433467. R.J.K. was supported by a fellowship within the Postdoc-Program of the German Academic Exchange Service (DAAD). S.U. acknowledges financial support from the Danish Council for Independent Research (Grant No. 4090-00125). K.K. and W.W. thank the Ohio Supercomputer Center for support under Project No. PAS0072. J.E.G. acknowledges the Camille and Henry Dreyfus Foundation for partial support.

ACKNOWLEDGMENT

We acknowledge the Analytical Spectroscopy Laboratory and the Surface Analysis Laboratory (NSF DMR-0114098) of The Ohio State University Department of Chemistry and Biochemistry and The Ohio State University Nanosystems Laboratory (NSL) and Center for Electron Microscopy and Analysis (CEMAS). The Advanced Light Source is supported by the Director, Office of Science, Office of Basic Energy Sciences, of the U.S. Department of Energy under Contract No. DE-AC02-05CH11231.

REFERENCES

(1) Xu, Y.; Yan, B.; Zhang, H.-J.; Wang, J.; Xu, G.; Tang, P.; Duan, W.; Zhang, S.-C. Large-Gap Quantum Spin Hall Insulators in Tin Films. *Phys. Rev. Lett.* **2013**, *111*, 136804.
(2) Novoselov, K. S.; Geim, A. K.; Morozov, S. V.; Jiang, D.; Katsnelson, M. I.; Grigorieva, I. V.; Dubonos, S. V.; Firsov, A. A. Two-Dimensional Gas of Massless Dirac Fermions in Graphene. *Nature* **2005**, *438*, 197-200.
(3) Novoselov, K. S.; Geim, A. K.; Morozov, S. V.; Jiang, D.; Zhang, Y.; Dubonos, S. V.; Grigorieva, I. V.; Firsov, A. A. Electric Field Effect in Atomically Thin Carbon Films. *Science* **2004**, *306*, 666-669.

(4) Xi, X.; Zhao, L.; Wang, Z.; Berger, H.; Forro, L.; Shan, J.; Mak, K. F. Strongly Enhanced Charge-Density-Wave Order in Monolayer $NbSe_2$. *Nat. Nanotechnol.* **2015**, *10*, 765-769.

(5) Jones, A. M.; Yu, H. Y.; Ghimire, N. J.; Wu, S. F.; Aivazian, G.; Ross, J. S.; Zhao, B.; Yan, J. Q.; Mandrus, D. G.; Xiao, D.; Yao, W.; Xu, X. D. Optical Generation of Excitonic Valley Coherence in Monolayer $WSe_2$. *Nat. Nanotechnol.* **2013**, *8*, 634-638.

(6) Radisavljevic, B.; Radenovic, A.; Brivio, J.; Giacometti, V.; Kis, A. Single-Layer $MoS_2$ Transistors. *Nat. Nanotechnol.* **2011**, *6*, 147-150.

(7) Butler, S. Z.; Hollen, S. M.; Cao, L.; Cui, Y.; Gupta, J. A.; Gutierrez, H. R.; Heinz, T. F.; Hong, S. S.; Huang, J.; Ismach, A. F. Progress, Challenges, and Opportunities in Two-Dimensional Materials Beyond Graphene. *ACS Nano* **2013**, *7*, 2898-2926.

(8) Zhang, Y.; Tan, Y.-W.; Stormer, H. L.; Kim, P. Experimental Observation of the Quantum Hall Effect and Berry's Phase in Graphene. *Nature* **2005**, *438*, 201-204.

(9) Ross, J. S.; Klement, P.; Jones, A. M.; Ghimire, N. J.; Yan, J.; Mandrus, D. G.; Taniguchi, T.; Watanabe, K.; Kitamura, K.; Yao, W.; Cobden, D. H.; Xu, X. Electrically Tunable Excitonic Light-Emitting Diodes Based on Monolayer $WSe_2$ P-N Junctions. *Nat. Nanotechnol.* **2014**, *9*, 268-272.

(10) Mak, K. F.; Lee, C.; Hone, J.; Shan, J.; Heinz, T. F. Atomically Thin $MoS_2$: A New Direct-Gap Semiconductor. *Phys. Rev. Lett.* **2010**, *105*, 136805.

(11) Splendiani, A.; Sun, L.; Zhang, Y.; Li, T.; Kim, J.; Chim, C.-Y.; Galli, G.; Wang, F. Emerging Photoluminescence in Monolayer $MoS_2$. *Nano Lett.* **2010**, *10*, 1271-1275.

(12) Zhao, W.; Ghorannevis, Z.; Chu, L.; Toh, M.; Kloc, C.; Tan, P.-H.; Eda, G. Evolution of Electronic Structure in Atomically Thin Sheets of $WS_2$ and $WSe_2$. *ACS Nano* **2012**, *7*, 791-797.

(13) Xiao, D.; Liu, G.-B.; Feng, W.; Xu, X.; Yao, W. Coupled Spin and Valley Physics in Monolayers of $MoS_2$ and Other Group-VI Dichalcogenides. *Phys. Rev. Lett.* **2012**, *108*, 196802.

(14) Mak, K. F.; McGill, K. L.; Park, J.; McEuen, P. L. The Valley Hall Effect in $MoS_2$ Transistors. *Science* **2014**, *344*, 1489-1492.

(15) Lee, J.; Mak, K. F.; Shan, J. Electrical Control of the Valley Hall Effect in Bilayer $MoS_2$ Transistors. *Nat. Nanotechnol.* **2016**, *11*, 421-425.

(16) Haugen, H.; Huertas-Hernando, D.; Brataas, A. Spin Transport in Proximity-Induced Ferromagnetic Graphene. *Phys. Rev. B* **2008**, *77*, 115406.

(17) Tombros, N.; Jozsa, C.; Popinciuc, M.; Jonkman, H. T.; van Wees, B. J. Electronic Spin Transport and Spin Precession in Single Graphene Layers at Room Temperature. *Nature* **2007**, *448*, 571-574.

(18) Wei, P.; Lee, S.; Lemaitre, F.; Pinel, L.; Cutaia, D.; Cha, W.; Katmis, F.; Zhu, Y.; Heiman, D.; Hone, J.; Moodera, J. S.; Chen, C.-T. Strong Interfacial Exchange Field in the Graphene/EuS Heterostructure. *Nat. Mater.* **2016,** in press. (doi:10.1038/nmat4603)

(19) Wang, Z.; Tang, C.; Sachs, R.; Barlas, Y.; Shi, J. Proximity-Induced Ferromagnetism in Graphene Revealed by the Anomalous Hall Effect. *Phys. Rev. Lett.* **2015**, *114*, 016603.

(20) Qian, X.; Liu, J.; Fu, L.; Li, J. Quantum Spin Hall Effect in Two-Dimensional Transition Metal Dichalcogenides. *Science* **2014**, *346*, 1344-1347.

(21) Kane, C. L.; Mele, E. J. Quantum Spin Hall Effect in Graphene. *Phys. Rev. Lett.* **2005**, *95*, 226801.

(22) Giovannetti, G.; Khomyakov, P. A.; Brocks, G.; Karpan, V. M.; van den Brink, J.; Kelly, P. J. Doping Graphene with Metal Contacts. *Phys. Rev. Lett.* **2008**, *101*, 026803.

(23) Wei, D.; Liu, Y.; Wang, Y.; Zhang, H.; Huang, L.; Yu, G. Synthesis of N-Doped Graphene by Chemical Vapor Deposition and Its Electrical Properties. *Nano Lett.* **2009**, *9*, 1752-1758.

(24) Kim, K. S.; Zhao, Y.; Jang, H.; Lee, S. Y.; Kim, J. M.; Kim, K. S.; Ahn, J.-H.; Kim, P.; Choi, J.-Y.; Hong, B. H. Large-Scale Pattern Growth of Graphene Films for Stretchable Transparent Electrodes. *Nature* **2009**, *457*, 706-710.

(25) Radisavljevic, B.; Kis, A. Mobility Engineering and a Metal-Insulator Transition in Monolayer $MoS_2$. *Nat. Mater.* **2013**, *12*, 815-820.

(26) Bostwick, A.; Ohta, T.; Seyller, T.; Horn, K.; Rotenberg, E. Quasiparticle Dynamics in Graphene. *Nat. Phys.* **2007**, *3*, 36-40.

(27) Ohta, T.; Bostwick, A.; Seyller, T.; Horn, K.; Rotenberg, E. Controlling the Electronic Structure of Bilayer Graphene. *Science* **2006**, *313*, 951-954.

(28) Livneh, T.; Sterer, E. Resonant Raman Scattering at Exciton States Tuned by Pressure and Temperature in $2H-MoS_2$. *Phys. Rev. B* **2010**, *81*, 195209.

(29) Katoch, J. Adatom-Induced Phenomena in Graphene. *Synth. Met.* **2015**, *210*, 68-79.

(30) Zhu, Z. Y.; Cheng, Y. C.; Schwingenschloegl, U. Giant Spin-Orbit-Induced Spin Splitting in Two-Dimensional Transition-Metal Dichalcogenide Semiconductors. *Phys. Rev. B* **2011**, *84*, 153402.

(31) Wu, S.; Ross, J. S.; Liu, G.-B.; Aivazian, G.; Jones, A.; Fei, Z.; Zhu, W.; Xiao, D.; Yao, W.; Cobden, D.; Xu, X. Electrical Tuning of Valley Magnetic Moment through Symmetry Control in Bilayer $MoS_2$. *Nat. Phys.* **2013**, *9*, 149-153.

(32) Baugher, B. W. H.; Churchill, H. O. H.; Yang, Y.; Jarillo-Herrero, P. Optoelectronic Devices Based on Electrically Tunable P-N Diodes in a Monolayer Dichalcogenide. *Nat. Nanotechnol.* **2014**, *9*, 262-267.

(33) Candini, A.; Klyatskaya, S.; Ruben, M.; Wernsdorfer, W.; Affronte, M. Graphene Spintronic Devices with Molecular Nanomagnets. *Nano Lett.* **2011**, *11*, 2634-2639.

(34) Zeng, H.; Dai, J.; Yao, W.; Xiao, D.; Cui, X. Valley Polarization in $MoS_2$ Monolayers by Optical Pumping. *Nat. Nanotechnol.* **2012**, *7*, 490-493.

(35) Lopez-Sanchez, O.; Lembke, D.; Kayci, M.; Radenovic, A.; Kis, A. Ultrasensitive Photodetectors Based on Monolayer $MoS_2$. *Nat. Nanotechnol.* **2013**, *8*, 497-501.

(36) Buscema, M.; Groenendijk, D. J.; Blanter, S. I.; Steele, G. A.; van der Zant, H. S. J.; Castellanos-Gomez, A. Fast and Broadband Photoresponse of Few-Layer Black Phosphorus Field-Effect Transistors. *Nano Lett.* **2014**, *14*, 3347-3352.

(37) Han, W.; Kawakami, R. K.; Gmitra, M.; Fabian, J. Graphene Spintronics. *Nat. Nanotechnol.* **2014**, *9*, 794-807.

(38) Jariwala, D.; Sangwan, V. K.; Lauhon, L. J.; Marks, T. J.; Hersam, M. C. Emerging Device Applications for Semiconducting Two-Dimensional Transition Metal Dichalcogenides. *ACS Nano* **2014**, *8*, 1102-1120.

(39) Eda, G.; Fanchini, G.; Chhowalla, M. Large-Area Ultrathin Films of Reduced Graphene Oxide as a Transparent and Flexible Electronic Material. *Nat. Nanotechnol.* **2008**, *3*, 270-274.

(40) Becerril, H. A.; Mao, J.; Liu, Z.; Stoltenberg, R. M.; Bao, Z.; Chen, Y. Evaluation of Solution-Processed Reduced Graphene Oxide Films as Transparent Conductors. *ACS Nano* **2008**, *2*, 463-470.

(41) Wang, Y.; Shao, Y.; Matson, D. W.; Li, J.; Lin, Y. Nitrogen-Doped Graphene and Its Application in Electrochemical Biosensing. *ACS Nano* **2010**, *4*, 1790-1798.

(42) Li, H.; Yin, Z.; He, Q.; Li, H.; Huang, X.; Lu, G.; Fam, D. W. H.; Tok, A. I. Y.; Zhang, Q.; Zhang, H. Fabrication of Single- and Multilayer $MoS_2$ Film-Based Field-Effect Transistors for Sensing No at Room Temperature. *Small* **2012**, *8*, 63-67.

(43) Late, D. J.; Huang, Y.-K.; Liu, B.; Acharya, J.; Shirodkar, S. N.; Luo, J.; Yan, A.; Charles, D.; Waghmare, U. V.; Dravid, V. P.; Rao, C. N. R. Sensing Behavior of Atomically Thin-Layered $MoS_2$ Transistors. *ACS Nano* **2013**, *7*, 4879-4891.

(44) Pumera, M. Graphene in Biosensing. *Mater. Today* **2011**, *14*, 308-315.

(45) Qu, L.; Liu, Y.; Baek, J.-B.; Dai, L. Nitrogen-Doped Graphene as Efficient Metal-Free Electrocatalyst for Oxygen Reduction in Fuel Cells. *ACS Nano* **2010**, *4*, 1321-1326.

(46) Jaramillo, T. F.; Jorgensen, K. P.; Bonde, J.; Nielsen, J. H.; Horch, S.; Chorkendorff, I. Identification of Active Edge Sites for Electrochemical $H_2$ Evolution from $MoS_2$ Nanocatalysts. *Science* **2007**, *317*, 100-102.

(47) Lukowski, M. A.; Daniel, A. S.; Meng, F.; Forticaux, A.; Li, L.; Jin, S. Enhanced Hydrogen Evolution Catalysis from Chemically Exfoliated Metallic $MoS_2$ Nanosheets. *J. Am. Chem. Soc.* **2013**, *135*, 10274-10277.

(48) Arguilla, M. Q.; Jiang, S.; Chitara, B.; Goldberger, J. E. Synthesis and Stability of Two-Dimensional Ge/Sn Graphane Alloys. *Chem. Mater.* **2014**, *26*, 6941-6946.

(49) Bianco, E.; Butler, S.; Jiang, S.; Restrepo, O. D.; Windl, W.; Goldberger, J. E. Stability and Exfoliation of Germanane: A Germanium Graphane Analogue. *ACS Nano* **2013**, *7*, 4414-4421.

(50) Jiang, S.; Bianco, E.; Goldberger, J. E. The Structure and Amorphization of Germanane. *J. Mater. Chem. C* **2014**, *2*, 3185-3188.

(51) Jiang, S.; Butler, S.; Bianco, E.; Restrepo, O. D.; Windl, W.; Goldberger, J. E. Improving the Stability and Optical Properties of Germanane *via* One-Step Covalent Methyl-Termination. *Nat. Comm.* **2014**, *5,* 3389.

(52) Lukatskaya, M. R.; Mashtalir, O.; Ren, C. E.; Dall'Agnese, Y.; Rozier, P.; Taberna, P. L.; Naguib, M.; Simon, P.; Barsoum, M. W.; Gogotsi, Y. Cation Intercalation and High Volumetric Capacitance of Two-Dimensional Titanium Carbide. *Science* **2013**, *341*, 1502-1505.

(53) Naguib, M.; Mashtalir, O.; Carle, J.; Presser, V.; Lu, J.; Hultman, L.; Gogotsi, Y.; Barsoum, M. W. Two-Dimensional Transition Metal Carbides. *ACS Nano* **2012**, *6*, 1322-1331.

(54) Khazaei, M.; Arai, M.; Sasaki, T.; Chung, C.-Y.; Venkataramanan, N. S.; Estili, M.; Sakka, Y.; Kawazoe, Y. Novel Electronic and Magnetic Properties of Two-Dimensional Transition Metal Carbides and Nitrides. *Adv. Funct. Mater.* **2013**, *23*, 2185-2192.

(55) Zhang, J.; Najmaei, S.; Lin, H.; Lou, J. $MoS_2$ Atomic Layers with Artificial Active Edge Sites as Transparent Counter Electrodes for Improved Performance of Dye-Sensitized Solar Cells. *Nanoscale* **2014**, *6*, 5279-5283.

(56) Yan, Z.; Jiang, C.; Pope, T.; Tsang, C.; Stickney, J.; Goli, P.; Renteria, J.; Salguero, T.; Balandin, A. Phonon and Thermal Properties of Exfoliated $TaSe_2$ Thin Films. *J. Appl. Phys.* **2013**, *114*, 204301.

(57) Fan, X.; Xu, P.; Li, Y. C.; Zhou, D.; Sun, Y.; Nguyen, M. A. T.; Terrones, M.; Mallouk, T. E. Controlled Exfoliation of $MoS_2$ Crystals into Trilayer Nanosheets. *J. Am. Chem. Soc.* **2016**, *138*, 5143-5149.

(58) Chen, K. P.; Chung, F. R.; Wang, M.; Koski, K. J. Dual Element Intercalation into 2D Layered $Bi_2Se_3$ Nanoribbons. *J. Am. Chem. Soc.* **2015**, *137*, 5431-5437.

(59) Asbrand, M.; Eisenmann, B.; Klein, J. Arsenidostannates with SnAs Nets Isostructural to Grey Arsenic - Synthesis and Crystal-Structure of $NaSn_2As_2$, $Na_{0.3}Sr_{0.7}Sn_2As_2$, $Na_{0.4}Sr_{0.6}Sn_2As_2$, $Na_{0.6}Ba_{0.4}Sn_2As_2$, and $K_{0.3}Sr_{0.7}Sn_2As_2$. *Z. Anorg. Allg. Chem.* **1995**, *621*, 576-582.

(60) Eisenmann, B.; Klein, J. Zintl-Phases with Layer Anions - Preparation and Crystal-Structures of the Isotypic Compounds $SrSn_2As_2$ and $Sr_{0.87}Ba_{0.13}Sn_2As_2$ and a Single-Crystal Structure Determination of Ksnsb. *Z. Anorg. Allg. Chem.* **1991**, *598*, 93-102.

(61) Eisenmann, B.; Rossler, U. Crystal Structure of Sodium Phosphidostannate(II), NaSnP. *Z. Kristallogr. - New Cryst. Struct.* **1998**, *213*, 28.

(62) Klein, J.; Eisenmann, B. Zintl Phases with Complex Anions - Synthesis and Structure-Analysis of Single-Crystals of KSnAs and $K_6Sn_3As_5$. *Mater. Res. Bull.* **1988**, *23*, 587-594.

(63) Evans, M. J.; Wu, Y.; Kranak, V. F.; Newman, N.; Reller, A.; Garcia-Garcia, F. J.; Haeussermann, U. Structural Properties and Superconductivity in the Ternary Intermetallic Compounds MAB (M=Ca, Sr, Ba; A=Al, Ga, In; B=Si, Ge, Sn). *Phys. Rev. B* **2009**, *80*, 064514.

(64) Lee, K.; Kaseman, D.; Sen, S.; Hung, I.; Gan, Z.; Gerke, B.; Poettgen, R.; Feygenson, M.; Neuefeind, J.; Lebedev, O. I.; Kovnir, K. Intricate Short-Range Ordering and Strongly Anisotropic Transport Properties of $Li_{1-x}Sn_{2+x}As_2$. *J. Am. Chem. Soc.* **2015**, *137*, 3622-3630.

(65) He, H.; Stearrett, R.; Nowak, E. R.; Bobev, S. $BaGa_2Pn_2$ (Pn= P, As): New Semiconducting Phosphides and Arsenides with Layered Structures. *Inorg. Chem.* **2010**, *49*, 7935-7940.

(66) Morar, J.; Wittmer, M. Metallic $CaSi_2$ Epitaxial Films on Si (111). *Phys. Rev. B* **1988**, *37*, 2618.

(67) Evans, M. J.; Lee, M. H.; Holland, G. P.; Daemen, L. L.; Sankey, O. F.; Häussermann, U. Vibrational Properties of the Gallium Monohydrides SrGaGeH, BaGaSiH, BaGaGeH, and BaGaSnH. *J. Solid State Chem.* **2009**, *182*, 2068-2073.

(68) Huang, H.; Liu, J.; Vanderbilt, D.; Duan, W. Topological Nodal-Line Semimetals in Alkaline-Earth Stannides, Germanides, and Silicides. *Phys. Rev. B* **2016**, *93*, 201114.

(69) Gibson, Q. D.; Schoop, L. M.; Muechler, L.; Xie, L. S.; Hirschberger, M.; Ong, N. P.; Car, R.; Cava, R. J. Three-Dimensional Dirac Semimetals: Design Principles and Predictions of New Materials. *Phys. Rev. B* **2015**, *91*, 205128.

(70) Manchon, A.; Koo, H.; Nitta, J.; Frolov, S.; Duine, R. New Perspectives for Rashba Spin-Orbit Coupling. *Nat. Mater.* **2015**, *14*, 871-882.

(71) Bjorling, T.; Noreus, D.; Haussermann, U. Polyanionic Hydrides from Polar Intermetallics $AeE_2$ (Ae = Ca, Sr, Ba; E = Al, Ga, In). *J. Am. Chem. Soc.* **2006**, *128*, 817-824.

(72) Evans, M. J.; Holland, G. P.; Garcia-Garcia, F. J.; Haeussermann, U. Polyanionic Gallium Hydrides from $AlB_2$-Type Precursors AeGaE (Ae = Ca, Sr, Ba; E = Si, Ge, Sn). *J. Am. Chem. Soc.* **2008**, *130*, 12139-12147.

(73) Jiang, S.; Arguilla, M. Q.; Cultrara, N. D.; Goldberger, J. E. Improved Topotactic Reactions for Maximizing Organic Coverage of Methyl Germanane. *Chem. Mater.* **2016**, *28*, 4735-4740.

(74) Morgan, W. E.; Van Wazer, J. R.; Stec, W. J. Inner-Orbital Photoelectron Spectroscopy of the Alkali Metal Halides, Perchlorates, Phosphates, and Pyrophosphates. *J. Am. Chem. Soc.* **1973**, *95*, 751-755.

(75) Okamoto, Y.; Carter, W. J.; Hercules, D. M. A Study of the Interaction of Pb-Sn Solder with $O_2$, $H_2O$, and $NO_2$ by X-Ray Photoelectron (ESCA) and Auger (AES) Spectroscopy. *Appl. Spectrosc.* **1979**, *33*, 287-293.

(76) Ball, D. W. The Structure and Vibrational Spectra of Isotopomers of SnOH and OSnH. *J. Mol. Struct.* **2003**, *626*, 217-221.

(77) Wang, X. F.; Andrews, L. Infrared Spectra of $M(OH)_{1,2,4}$ (M = Pb, Sn) in Solid Argon. *J. Phys. Chem. A* **2005**, *109*, 9013-9020.

(78) Myneni, S. C. B.; Traina, S. J.; Waychunas, G. A.; Logan, T. J. Vibrational Spectroscopy of Functional Group Chemistry and Arsenate Coordination in Ettringite. *Geochim. Cosmochim. Acta* **1998**, *62*, 3499-3514.

(79) Grimme, S. Semiempirical GGA-Type Density Functional Constructed with a Long-Range Dispersion Correction. *J. Comput. Chem.* **2006**, *27*, 1787-1799.

(80) Cao, Y.; Mishchenko, A.; Yu, G. L.; Khestanova, E.; Rooney, A. P.; Prestat, E.; Kretinin, A. V.; Blake, P.; Shalom, M. B.; Woods, C.; Chapman, J.; Balakrishnan, G.; Grigorieva, I. V.; Novoselov, K. S.; Piot, B. A.; Potemski, M.; Watanabe, K.; Taniguchi, T.; Haigh, S. J.; Geim, A. K.; Gorbachev, R. V. Quality Heterostructures from Two-Dimensional Crystals Unstable in Air by Their Assembly in Inert Atmosphere. *Nano Lett.* **2015**, *15*, 4914-4921.

(81) Wood, J. D.; Wells, S. A.; Jariwala, D.; Chen, K. S.; Cho, E.; Sangwan, V. K.; Liu, X. L.; Lauhon, L. J.; Marks, T. J.; Hersam, M. C. Effective Passivation of Exfoliated Black Phosphorus Transistors against Ambient Degradation. *Nano Lett.* **2014**, *14*, 6964-6970.

(82) Pei, J. J.; Gai, X.; Yang, J.; Wang, X. B.; Yu, Z. F.; Choi, D. Y.; Luther-Davies, B.; Lu, Y. R. Producing Air-Stable Monolayers of Phosphorene and Their Defect Engineering. *Nat. Comm.* **2016**, *7*, 10450.

(83) Ryder, C. R.; Wood, J. D.; Wells, S. A.; Yang, Y.; Jariwala, D.; Marks, T. J.; Schatz, G. C.; Hersam, M. C. Covalent Functionalization and Passivation of Exfoliated Black Phosphorus *via* Aryl Diazonium Chemistry. *Nat. Chem.* **2016**, *8*, 598-603.

(84) Ooi, N.; Rairkar, A.; Adams, J. B. Density Functional Study of Graphite Bulk and Surface Properties. *Carbon* **2006**, *44*, 231-242.

(85) Koren, E.; Lortscher, E.; Rawlings, C.; Knoll, A. W.; Duerig, U. Adhesion and Friction in Mesoscopic Graphite Contacts. *Science* **2015**, *348*, 679-683.

(86) Levita, G.; Molinari, E.; Polcar, T.; Righi, M. C. First-Principles Comparative Study on the Interlayer Adhesion and Shear Strength of Transition-Metal Dichalcogenides and Graphene. *Phys. Rev. B* **2015**, *92,* 085434.

**Table of Contents Image**

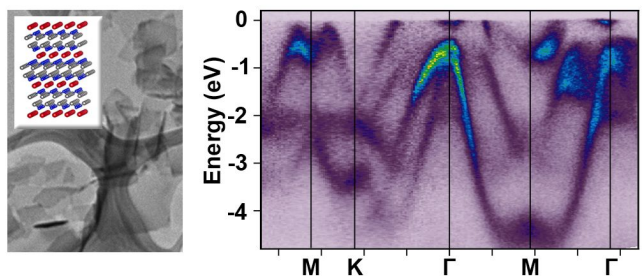